\documentclass[prl,aps,amssymb,twocolumn,showpacs,floatfix]{revtex4}

\usepackage{graphicx}

\begin{document}

\title{Reduced Visibility of Rabi Oscillations in Superconducting Qubits}

\author{Florian Meier$^{1,2}$}\email{meier@physics.ucsb.edu}
\author{Daniel Loss$^2$}
\affiliation{$^1$Center for Spintronics and Quantum Computation,
University of California, Santa Barbara, California 93106, USA
\\
$^2$Department of Physics and Astronomy, University of Basel,
Klingelbergstrasse 82, 4056 Basel, Switzerland }
\date{\today}

\begin{abstract}
Coherent Rabi oscillations between quantum states of
superconducting micro-circuits have been observed in a number of
experiments, albeit with a visibility which is typically much
smaller than unity. Here, we show that the coherent coupling to
background charge fluctuators [R.~W. Simmonds {\it et al.}, Phys.
Rev. Lett. {\bf 93}, 077003 (2004)] leads to a significantly
reduced visibility if the Rabi frequency is comparable to the
coupling energy of micro-circuit and fluctuator. For larger Rabi
frequencies, transitions to the second excited state of the
superconducting micro-circuit become dominant in suppressing the
Rabi oscillation visibility. We also calculate the probability for
Bogoliubov quasi-particle excitations in typical Rabi oscillation
experiments.
\end{abstract}


\maketitle

Possible applications in quantum information processing have
renewed the interest in quantum coherent phenomena of
micrometer-scale Josephson junction (JJ)
circuits~\cite{makhlin:01}. During the past years, several
experiments have demonstrated Rabi oscillations of a macroscopic
variable including charge~\cite{nakamura:99},
phase~\cite{martinis:02}, a combinations of
both~\cite{vion:02,yu:02}, and flux~\cite{chiorescu:03}, which
persist up to several microseconds~\cite{vion:02}. However, in
many of the reported experiments, the Rabi oscillation visibility
is significantly smaller than unity even at times short compared
to the decoherence time. Characteristic values are as small as
$10$\% (Ref.~\onlinecite{vion:02}) and $50$\%
(Refs.~\onlinecite{martinis:02,simmonds:04}), respectively, which
indicates either substantial leakage out of the computational
basis or a pronounced randomization of the state occupation over
short time-scales. In the following, we discuss three possible
mechanisms that result in a reduced visibility, namely (i)
coherent coupling to a background fluctuator which induces
transitions between the eigenstates of the superconducting
micro-circuit~\cite{simmonds:04}; (ii) population of higher
excited states caused by a strong driving field, i.e., the AC
current or voltage applied to drive Rabi oscillations; and (iii)
the excitation of Bogoliubov quasi-particles.

Background charge fluctuators are presently considered one of the
dominant source of decoherence in superconducting qubits
(squbits)~\cite{nakamura:02,paladino:02,martinis:03,galperin:03}.
However, the theoretical analysis of fluctuators has so far been
focused on fluctuators with an incoherent dynamics which are
coupled to the squbit by an Ising-like interaction. Motivated by
the results of Refs.~\onlinecite{simmonds:04} and
\onlinecite{cooper:04}, we show that the coherent dynamics of a
squbit-fluctuator system with a transverse exchange coupling leads
to a reduced visibility and beating in the Rabi oscillation signal
of the squbit. For Rabi oscillation frequencies large compared to
the squbit-fluctuator coupling strength $J$, the fluctuator does
not provide an efficient mechanism for visibility reduction on
short time-scales. We show that, in this regime, coupling to the
second excited squbit state suppresses the visibility to $0.7$ for
Rabi frequencies larger than $150 \,~{\rm MHz}$ in the experiment
in Ref.~\onlinecite{simmonds:04}. This mechanism persists for
adiabatic switching, in stark contrast to the leakage scenarios
discussed previously~\cite{fazio:99,steffen:03}. Finally, we
calculate the excitation rate of Bogoliubov quasi-particles and
show that the quasi-particle excitation probability is small as
long as the JJ is in a zero-voltage state.

{\it Coherent squbit-fluctuator coupling. --} While fluctuators,
trapped charges in bi-stable potentials, have so far mainly been
considered as source of decoherence for squbits, recent
experiments provide strong evidence that some fluctuators are
coupled coherently to a squbit. An important conclusion drawn from
the anti-crossings in the microwave spectra of a
phase-squbit~\cite{simmonds:04} and the subsequent observation of
real-time oscillations~\cite{cooper:04} is that the
squbit-fluctuator coupling has a transverse component rather than
the Ising-form considered in previous theoretical
analysis~\cite{paladino:02,galperin:03}. Because of the transverse
coupling component, fluctuators induce {\it transitions} between
different squbit basis states, randomizing the occupation
probabilities. In order to analyze the dynamics of a coupled
squbit-fluctuator system, we introduce a pseudo-spin notation and
define $\hat{s}_z = (|1\rangle \langle 1|-|0\rangle\langle 0|)/2$
and $\hat{s}_x = (|1\rangle \langle 0| + |0\rangle \langle 1|)/2$
in terms of the ground (first excited) state $|0\rangle$
($|1\rangle$) of the squbit, such that $|s_z=\downarrow\rangle$
represents the squbit ground state. Similarly, we define the
pseudo-spin states $|I_z = \downarrow \rangle$ and $|I_z =
\uparrow \rangle $ as the ground and first excited states of the
fluctuator, respectively. Rabi oscillations between the squbit
basis states are driven by an AC current in resonance with the
level splitting $\hbar \omega_{10}$ of the squbit. In our
pseudo-spin notation, the AC current acts as transverse magnetic
field $b_x$ and, in the co-rotating frame,
\begin{equation}
\hat{H} = \delta \hat{I}_z + J\left(\hat{s}_x \hat{I}_x +
\hat{s}_y \hat{I}_y \right) + b_x  \hat{s}_x, \label{eq:ham}
\end{equation}
where $\delta = \hbar(\omega_{eg} - \omega_{10})$ is the detuning
of the fluctuator level splitting $\hbar \omega_{eg}$ relative to
the squbit. The ansatz for the transverse squbit-fluctuator
exchange coupling with coupling strength $J$ is microscopically
motivated from the observation that the critical current of a JJ
may depend on the position of a fluctuator in its bi-stable
potential~\cite{simmonds:04}. As shown below, a fluctuator
influences the squbit dynamics strongly if $|\delta|<J$ or
$|\delta \pm b_x|<J$. For the JJ in Ref.~\onlinecite{simmonds:04},
$J/h \simeq 25~{\rm Mhz}$ while the typical level spacing between
different fluctuator resonances is of order $60$~MHz. This allows
us to restrict our analysis to one fluctuator with minimum
$|\delta|$ or $|\delta \pm b_x|$ first. For the experimental
temperature $T \simeq 20 \, {\rm mK} \ll \hbar \omega_{10}/k_B,
\hbar \omega_{eg}/k_B$, at the beginning of the Rabi pulse, $t=0$,
both squbit and fluctuator are in their ground-state and
$|\psi(0)\rangle = |\! \downarrow; \downarrow\rangle$ in the
product basis $|s_z;I_z\rangle$ of the squbit-fluctuator system.
The experimentally accessible quantity is the probability $p_1(t)$
for the squbit to occupy its first excited state as a function of
Rabi pulse duration, $p_1(t) = \sum_{I_z=\uparrow,\downarrow}
|\langle \uparrow; I_z | \psi(t) \rangle|^2$. While the energy
level splitting $\hbar \omega_{eg}$ of a fluctuator is fixed, the
squbit level splitting $\hbar \omega_{10}$ can be tuned via the DC
bias current through the JJ, which allows one to measure squbit
Rabi oscillations for varying $\delta$ and $b_x$.

We now calculate the dynamics of $|\psi(0)\rangle=|\!\downarrow
;\downarrow \rangle$ and $p_1(t)$ as a function of $\delta$ and
$b_x$. The squbit-fluctuator exchange coupling gives rise to a
linear dependence of the eigenenergies on $J$ for $|\delta|
\lesssim J$ and $|\delta \pm b_x| \lesssim J$, where
cross-relaxation processes between squbit and fluctuator reduce
the visibility of the squbit Rabi oscillations~\cite{rem1}. While
$p_1(t)$ is readily obtained from integration of the Schr\"odinger
equation for arbitrary $b_x$, $\delta$, and $J$, here we focus on
the two cases $\delta = 0$ and $\delta \pm b_x = 0$, where the
cross-relaxation between squbit and fluctuator is most efficient.
For $\delta=0$, we find
\begin{eqnarray}
\left. p_1(t) \right|_{\delta=0} =&& \frac{1}{2} \bigl[ 1 - \cos
\bigl(\frac{J t}{2 \hbar}\bigr) \cos \bigl(
\frac{\sqrt{J^2 + 4 b_x^2} t}{2 \hbar}\bigr) \label{eq:pdyn-1} \\
 && - \frac{\sin (\frac{J t}{2}) \sin
(\frac{\sqrt{J^2 + 4 b_x^2} t}{2 \hbar})}{\sqrt{1+(2b_x/J)^2}}
\bigr] \nonumber \\ \stackrel{|b_x|/J \gg 1}{\longrightarrow} &&
\frac{1}{2} \bigl[ 1 - \cos \bigl( \frac{J t}{2 \hbar}\bigr) \cos
\bigl( \frac{b_x t}{\hbar} \bigr) \bigr]. \nonumber
\end{eqnarray}
For $\delta \pm b_x=0$, the exact expression for $p_1(t)$ is too
long to be presented here, but for $|b_x|/J \gtrsim 2$ it is well
approximated by
\begin{equation}
\left. p_1 (t)\right|_{\delta \pm b_x =0}  \simeq \frac{1}{2}
\bigl[1 - \cos \bigl( \frac{J t}{4 \hbar} \bigr) \, \cos \bigl(
\frac{b_x t}{\hbar}\bigr) \bigr]. \label{eq:pdyn-2}
\end{equation}
Equations~(\ref{eq:pdyn-1}) and (\ref{eq:pdyn-2}) describe the
Rabi oscillations of a squbit in the presence of a fluctuator. The
transverse coupling to a fluctuator introduces an additional
Fourier component which leads to beating of the Rabi-oscillation
signal, in agreement with experimental results, where typically
$|b_x|/J \geq 2$. For short time-scales, the $J$-dependent factor
in Eqs.~(\ref{eq:pdyn-1}) and (\ref{eq:pdyn-2}) leads to a
decrease in Rabi oscillation visibility. In particular, the first
maximum in $p_1(t)$ is reduced relative to unity.  For $|b_x|/J
\gtrsim 2$, we obtain $p_1(t = \pi \hbar/b_x) \simeq 1 - (\pi J/4
b_x)^2$ and $p_1(t = \pi \hbar/b_x) \simeq 1 - (\pi J/8 b_x)^2$
for $\delta = 0$ and $\delta \pm b_x=0$, respectively~\cite{rem2}.
While a resonant fluctuator reduces $p_1(t= \pi \hbar/b_x)$ to
$0.85$ for $|b_x|/J \simeq 2$, Rabi oscillations with amplitude
$1$ are predicted to emerge for $|b_x|/J \gg 1$, in stark contrast
to experimental results where the first local maximum of $p_1(t)$
is of order $0.5$ even for large $b_x$~\cite{simmonds:04}.

\begin{figure}
\centerline{\mbox{\includegraphics[width=9.5cm]{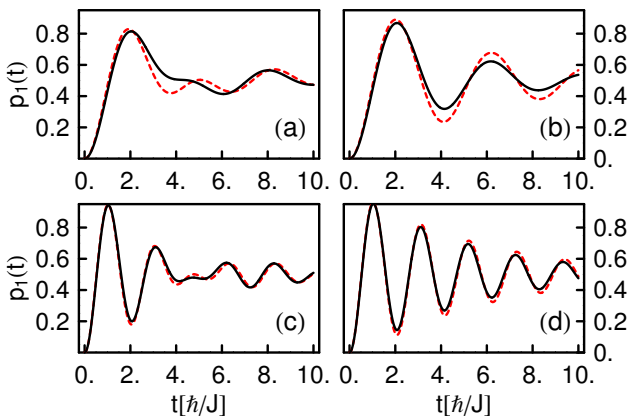}}}
\caption{(color online). Rabi oscillations for a squbit-fluctuator
system. The probability $p_1(t)$ to find the squbit in state
$|1\rangle$ is obtained from numerical integration of
Eq.~(\ref{eq:mastereq}) (solid line) and the analytical solution
Eq.~(\ref{eq:pdyn-3}) (dashed line) which is valid for $b_x/J \gg
1$. For weak decoherence of the fluctuator, $\gamma/J < 1$ [(a)
and (c)], $p_1(t)$ shows damped beating. For $\gamma/J > 1$,
damped single-frequency oscillations are restored. The fluctuator
leads to a reduction of the first maximum in $p_1(t)$ to $\sim
0.8$ [(a) and (b)] and $\sim 0.9$ [(c) and (d)], respectively. The
parameters are (a) $b_x/J=1.5$, $\gamma/J=0.5$; (b) $b_x/J=1.5$,
$\gamma/J=1.5$; (c) $b_x/J=3$, $\gamma/J=0.5$; (d) $b_x/J=3$,
$\gamma/J=1.5$.}\label{Fig1}
\end{figure}

In a realistic system, decoherence of the squbit and fluctuator
dynamics will damp out both the Rabi oscillations and the beating
predicted for $p_1(t)$ in Eqs.~(\ref{eq:pdyn-1}) and
(\ref{eq:pdyn-2}). In order to discuss reduced visibility in the
presence of decoherence, we focus on a simple model system in
which the fluctuator has a finite decoherence rate $\gamma/\hbar$
while its relaxation rate vanishes (`diagonal coupling' of a bath
and fluctuator). The dynamics of the system is determined by the
Master equation for the density matrix~\cite{blum:96},
\begin{equation}
\dot{\hat{\rho}}(t) = - \frac{i}{\hbar} [\hat{H}, \hat{\rho}(t)]
 + (\gamma/2 \hbar) \left( 4 \hat{I}_z \hat{\rho}(t)  \hat{I}_z -
 \hat{\rho}(t) \right).
\label{eq:mastereq}
\end{equation}
A similar system was analyzed in Ref.~\onlinecite{gassmann:02}
using different numerical and approximate analytical approaches.
Here, we provide an analytical solution of Eq.~(\ref{eq:mastereq})
for $|b_x|/J \gg 1$ where, to leading order in $J$, $\hat{H}
\simeq \delta \hat{I}_z+ J \hat{s}_x \hat{I}_x + b_x
 \hat{s}_x $. Then, the sixteen
coupled differential equations in Eq.~(\ref{eq:mastereq}) decouple
into sets of four differential equations. We calculate $p_1(t) =
\sum_{I_z = \uparrow, \downarrow} \langle
\uparrow;I_z|\hat{\rho}(t)|\uparrow;I_z\rangle$ from the explicit
solution of Eq.~(\ref{eq:mastereq}). For $\delta = 0$ and
$\hat{\rho}(0)=| \! \downarrow; \downarrow \rangle \langle
\downarrow; \downarrow \! |$,
\begin{eqnarray} p_1(t) &&\simeq
\frac{1}{2} - \frac{1}{4} \Re \Bigl\{ \bigl[ \bigl( 1 -
\frac{\gamma}{\sqrt{\gamma^2 - J^2}}\bigr) e^{- ( \gamma +
\sqrt{\gamma^2 - J^2}) t/2 \hbar} \hspace*{0.5cm} \nonumber \\ &&+
\bigl( 1 + \frac{\gamma}{\sqrt{\gamma^2 - J^2}}\bigr) e^{- (
\gamma - \sqrt{\gamma^2 - J^2}) t/2 \hbar} \bigr] e^{- i b_x
t/\hbar} \Bigr\}. \label{eq:pdyn-3}
\end{eqnarray}
Note the intriguing dependence of $p_1(t)$ on the fluctuator
decoherence rate $\gamma/\hbar$ and the coupling strength $J$
(Fig.~\ref{Fig1}). For $\gamma/J \ll 1$,
\begin{equation}
p_1(t) \simeq \frac{1}{2} \bigl[ 1 - e^{-\gamma t/2 \hbar} \cos
\bigl(\frac{J t}{2 \hbar}\bigr) \, \cos \bigl( \frac{b_x t}{\hbar}
\bigr) \bigr] \label{eq:pdyn-4}
\end{equation}
shows the beating already derived in Eq.~(\ref{eq:pdyn-1}), but
with a finite damping rate $\gamma/2 \hbar$. In contrast, for
$\gamma/J \gg 1$,
\begin{equation}
p_1(t) \simeq \frac{1}{2} \bigl[ 1 - e^{- 4 J^2  t/\hbar \gamma }
\cos \bigl( \frac{b_x t}{\hbar} \bigr) \bigr] \label{eq:pdyn-5}
\end{equation}
exhibits single-frequency oscillations with a damping rate $4
J^2/\hbar \gamma$ (see also Ref.~\onlinecite{gassmann:02}). This
can be understood from the observation that, for $\gamma \gg J$,
the phase of the fluctuator is randomized on a time-scale short
compared to $h/J$ and the mean field $J \langle \hat{I}_x\rangle$
acting on the squbit is averaged out to leading order, which
restores the single-frequency oscillations in
Eq.~(\ref{eq:pdyn-5}).

We discuss next to which extent far off-resonant fluctuators with
$|\delta| \gg |b_x| \gg J$ reduce the Rabi oscillation visibility.
Because of the perturbative parameter $J /|\delta| \ll 1$, the
visibility reduction in $p_1(t) \simeq (1-J^2/4\delta^2) \sin^2
(b_x t/2\hbar)$ is small for a single off-resonant fluctuator.
More generally, a large number of off-resonant fluctuators
interacts with the squbit, and the problem is closely related to
the dynamics of an electron spin coupled to a bath of nuclear
spins by the hyperfine contact interaction~\cite{khaetskii:02}. To
leading order in the coupling constants, a set $\{i\}$ of
off-resonant fluctuators with coupling constants $J_i$ and
detunings $\delta_i$ reduces the Rabi oscillation visibility by
$\sum_{i} J_i^2/4\delta_i^2$. Introducing the distribution
function $P(E)$ of the fluctuator level splittings, for constant
$J_i=J$,
\begin{equation}
\sum_{i} J_i^2/4\delta_i^2 = \frac{J^2}{4} \int_{|E-\hbar
\omega_{10}| > |b_x|; E>E_c} \frac{dE \, P(E)}{(\hbar
\omega_{10}-E)^2}, \label{eq:or-red}
\end{equation}
where the integral is evaluated for all energies with $|E-\hbar
\omega_{10}| > |b_x|$ (off-resonance condition) larger than a
lower energy cut-off $E_c$ given by the fluctuator decoherence
rate. Evaluating the integral for the distribution function $P(E)
\propto 1/E$ characteristic for fluctuator level spacings
($1/f$-noise), we find $\sum_{i} J^2/4\delta_i^2 \simeq 2 J^2
P(\hbar \omega_{10})/4 |b_x| = 0.2 J/|b_x|$ for the experimental
parameters of Ref.~\onlinecite{simmonds:04}. If cross correlations
between fluctuators can be neglected, the reduction of the first
maximum in $p_1(t)$ is obtained by summing the contributions from
resonant fluctuators with $\delta
 \simeq 0$ and $\delta \pm b_x \simeq 0$ [Eqs. (\ref{eq:pdyn-1}) and
(\ref{eq:pdyn-2})] and off-resonant fluctuators
[Eq.~(\ref{eq:or-red})]. An upper bound is given by $1-p_1(t = \pi
\hbar/b_x)\leq 2 J^2 P(\hbar \omega_{10})/4 |b_x| + (3/32)(\pi
J/b_x)^2$, which decreases with increasing $|b_x|$. This shows
that, even when both resonant and off-resonant fluctuators are
taken into account, the small visibility $p_1(t= \pi \hbar/b_x)
\leq 0.5$ observed in the limit of large Rabi frequencies,
$|b_x|/h \gg J/h$ in Ref.~\onlinecite{simmonds:04} cannot be
effected by fluctuators.

\begin{figure}
\centerline{\mbox{\includegraphics[width=8.5cm]{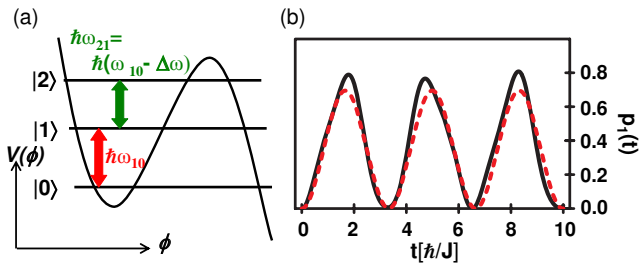}}}
\caption{(color online). (a) Level scheme for the squbit in
Refs.~\onlinecite{martinis:03,simmonds:04}. (b) $p_1(t)$ obtained
from numerical integration of the Schr\"odinger equation (solid
line) in comparison with the analytical result in
Eq.~(\ref{eq:pdyn-6}) (dashed line) for $b_x/\hbar \Delta \omega =
1/3$.}\label{Fig2}
\end{figure}

{\it Energy shifts induced by AC driving field. --} Exploring the
visibility reduction at a time-scale of $10\, {\rm ns}$ requires
Rabi frequencies $|b_x|/h \gtrsim 100 \,{\rm MHz}$. We show next
that, in this regime, transitions to the second excited squbit
state lead to an oscillatory behavior in $p_1(t)$ with a
visibility smaller than $0.7$. For characteristic parameters of a
phase-squbit, the second excited state $|2\rangle$ is
energetically separated from $|1\rangle$ by $\omega_{21} = 0.97
\omega_{10}$~\cite{martinis:03}. Similarly to $|0\rangle$ and
$|1\rangle$, the state $|2\rangle$ is localized around the local
energy minimum in Fig.~\ref{Fig2}(a). For adiabatic switching of
the AC current, transitions to $|2\rangle$ can be neglected as
long as $|b_x| \ll \hbar \Delta \omega = \hbar (\omega_{10} -
\omega_{21}) \simeq 0.03 \hbar \omega_{10}$.  However, for $b_x$
comparable to $\hbar \Delta \omega$, the applied AC current
strongly couples $|1\rangle$ and $|2\rangle$ because $\langle
2|\hat{\phi}|1\rangle \neq 0$, where $\hat{\phi}$ is the phase
operator. For typical parameters, $b_x/\hbar \Delta \omega$ ranges
from $0.05$ to $1$, depending on the irradiated
power~\cite{martinis:03,simmonds:04}. Taking into account the
second excited state of the phase-squbit, the squbit Hamiltonian
in the rotating frame is~\cite{steffen:03}
\begin{equation}
\hat{H} = - \hbar \Delta \omega |2\rangle \langle 2| + b_x \left(
|0\rangle \langle 1| + \sqrt {2} |1\rangle \langle 2| + H.c.
\right)/2. \label{eq:3state}
\end{equation}
In the following, we neglect fluctuators, which is valid for the
short-time dynamics if $|b_x| \gg J$.

The time-evolution of $|\psi(0)\rangle = |0\rangle$ is readily
calculated by integration of the Schr\"odinger equation. Expanding
$|\psi(t)\rangle$ in $(b_x/2 \hbar \Delta \omega)$, we find that
\begin{equation}
p_1 (t) = \left[1 - \frac{3}{2}\Bigl(\frac{b_x}{2\hbar \Delta
\omega} \Bigr)^2 \right]^2 \sin^2 \frac{b_x \bigl[1- (b_x/2\hbar
\Delta \omega)^2\bigr] t}{\hbar} \label{eq:pdyn-6}
\end{equation}
exhibits single-frequency oscillations with reduced visibility
[Fig.~\ref{Fig2}(b)]. Part of the visibility reduction can be
traced back to leakage into state $|2\rangle$.  More subtly,
off-resonant transitions from $|1\rangle$ to $|2\rangle$ induced
by the driving field lead to an energy shift of $|1\rangle$, such
that the transition from $|0\rangle$ to $|1\rangle$ is no longer
resonant with the driving field, which also reduces the
visibility. For $b_x/2 \hbar \Delta \omega = 1/3$, corresponding
to $b_x/h = 150~\, {\rm MHz}$ in Ref.~\onlinecite{simmonds:04},
Eq.~(\ref{eq:pdyn-6}) predicts a visibility of $0.7$. Note that
the off-resonant transition to $|2\rangle$ discussed here is
induced by a large {\it amplitude} of the driving field and not by
additional {\it frequency} components that result, e.g., from
non-adiabatic switching of the driving field~\cite{fazio:99} or
current noise in the micro-circuit~\cite{burkard:04}. Rather, the
visibility reduction in Eq.~(\ref{eq:pdyn-6}) corresponds to a
steady-state solution of the Schr\"odinger equation and persists
even for adiabatic switching~\cite{rem3}. Whether such
off-resonant transitions are responsible for the reduced Rabi
oscillation visibility can be tested experimentally from the
dependence of the Rabi frequency and amplitude on $b_x$.
Equation~(\ref{eq:pdyn-6}) predicts a non-linear dependence of the
Rabi frequency on $b_x$, while the visibility is predicted to
decrease quadratically with $b_x$. Decreasing visibility with
increasing driving field has indeed been observed in some Rabi
oscillation experiments~\cite{vion:03}, but further quantitative
analysis is required to determine the explicit functional
dependence.

{\it Excitation of quasi-particles.--} Transient high-frequency
components in the switching pulses induce leakage to states
outside the computational basis. Bogoliubov quasi-particle
excitations represent a large class of excited states which is
often ignored for the discussion of leakage because the excitation
gap is large. For both charge-based and phase-based squbits, the
excitation of quasi-particles during the measurement process has
been quantified and is known to limit the decoherence
time~\cite{nakamura:99,choi:01,lang:03}. In phase-squbits,
quasi-particles trigger leakage to finite-voltage states or to the
excited squbit-states $|2\rangle$ or $|3\rangle$ by incoherent
tunneling across the JJ at a rate determined by the
single-particle tunneling rate and the total number of
quasi-particles~\cite{lang:03}.

While the number of quasi-particles created during read-out can be
decreased by increasing experimental waiting times~\cite{lang:03},
Bogoliubov quasi-particles are also excited {\it during} an
experimental cycle by high-frequency Fourier components of an
external current or voltage pulse with $\omega
> 2\Delta/\hbar$, where $\Delta$ is the energy gap of the
superconductor. Microscopic mechanisms for quasi-particle
excitation include, e.g., dissipative tunneling through the JJ. We
calculate the excitation rate $\Gamma$ of quasi-particle pairs for
an oscillating current component $\delta I_\omega \cos (\omega t)$
with $\omega > 2 \Delta/\hbar$. Using a semiclassical
approximation, the high-frequency current induces oscillations
$\delta \phi (t)$ of the superconducting phase around its
equilibrium value $\phi_0 = \arcsin (I/I_c) \simeq \pi/2$, where
$I$ and $I_c$ are the DC bias current and critical current of the
JJ, respectively. Amplitude and frequency of $\delta \phi(t)$ are
determined by the externally imposed current, $ \delta \phi (t) =
- [(e \delta I_\omega)/(\hbar \omega^2 C)] \cos \omega t$, where
$C$ is the JJ capacitance. Dissipative tunneling through the JJ --
processes in which a quasi-particle pair is excited under
absorption of energy $\hbar \omega$ from the bias current -- is
described by both the normal current and the Josephson
cosine-term~\cite{harris:75}. To lowest non-vanishing order in $(e
\delta I_\omega)/(\hbar \omega^2 C)$, the time-averaged power
dissipation $\bar{P}$ at the JJ is~\cite{harris:75}
\begin{eqnarray}
\bar{P} &=& \frac{1}{2} \left(\frac{\delta I_\omega}{C \omega}
\right)^2 \frac{e}{\hbar \omega} \left[I_n(\hbar \omega) +
I_2(\hbar \omega) \cos \phi_0 \right] \nonumber \\ & \simeq  &
\frac{1}{2} \left(\frac{\delta I_\omega}{C \omega} \right)^2
\frac{e}{\hbar \omega} I_n(\hbar \omega) , \label{eq:dissipation}
\end{eqnarray}
with the standard expressions for the normal current $I_n$ and the
Josephson cosine-term $I_2$. For $\hbar \omega \rightarrow 2
\Delta^+$, $I_n$ is determined by the normal-state resistance
$R_n$ of the JJ~\cite{ambegaokar:63}, $I_n (2 \Delta^+) =
(\pi/4R_n )(2\Delta/e)$. The second line of
Eq.~(\ref{eq:dissipation}) is valid for squbits biased close to
the critical current, where $\phi_0 \simeq \pi/2$. The
quasi-particle excitation rate is obtained from $\bar{P}$
via~\cite{rem4}
\begin{equation}
\Gamma = \frac{\bar{P}}{\hbar \omega} = \frac{\pi\delta
I_\omega^2}{8\hbar C^2 R_n\omega^3} . \label{eq:qprate}
\end{equation}
$\Gamma$ is negligibly small unless $C \leq 1 \, {\rm fF}$ and
$\delta I_\omega \geq 1 \, {\rm pA}$. With typical parameters $R_n
= 29 \, \Omega$, $C = 1 \, {\rm fF}$, and $\omega = 4 \, {\rm K}
k_B/\hbar$,  we obtain $\Gamma=1 \,{\rm msec}^{-1} \times (\delta
I_\omega/{\rm pA})^2$. For instantaneous switching, a current
pulse with amplitude $I_{ac} = 10 \, {\rm nA}$ and duration $T=10
\, {\rm nsec}$, $\delta I_\omega \simeq I_{ac}/\omega T = 1 {\rm
pA}$ for $\omega = 4 \, {\rm K} k_B/\hbar$. These values show that
quasi-particle excitation by dissipative tunneling cannot explain
the substantial visibility reduction evidenced in current Rabi
oscillation experiments for phase squbits. However, the excitation
of Bogoliubov quasi-particles is relevant in view of the ultimate
goal of reducing quantum gate errors to less than $10^{-4}$.

{\it Acknowledgments. --} We acknowledge helpful discussions with
B.L. Altshuler, G. Burkard, H. Gassmann, F. Marquardt, J.M.
Martinis, and M. Steffen. This work was supported by ONR, DARPA,
the Swiss NCCR Nanoscience, and the Swiss NSF.

\bibliographystyle{prsty}

\begin{thebibliography}{10}
\bibitem{rem1}
In the opposite regime where $|\delta|,|\delta \pm b_x| \gg J$,
the
  fluctuator-subit coupling can be treated perturbatively (see below).
\bibitem{rem2}
In the following, we will quantify the visibility by $p_1(t=\pi
\hbar/b_x)$.
\bibitem{rem3}
The coupling to $|2\rangle$ also reduces the visibility in
free-induction decay
  experiments because it implies a reduced fidelity for the initial
  $\pi/2$-pulse.
\bibitem{rem4}
This result can also be derived from Fermi's golden rule if
electron tunneling
  across the JJ and the periodic phase variation $\delta \phi(t)$ are treated
  as perturbations.
\end{thebibliography}

\end{document}